\documentclass[%
prb
preprint%
twocolumn,amsmath,amssymb,%
reprint,%
superscriptaddress,
]{revtex4-1}

\usepackage{graphicx,verbatim}
\usepackage{dcolumn}
\usepackage{sidecap}
\begin{document}

\title{Effect of nonadiabatic electronic-vibrational interactions on the transport properties of single-molecule junctions}

\author{A.\ Erpenbeck}
\affiliation{Institut f\"ur Theoretische Physik und Interdisziplin\"ares Zentrum f\"ur Molekulare Materialien, Friedrich-Alexander-Universit\"at Erlangen-N\"urnberg, Staudtstr.\ 7/B2, D-91058 Erlangen, Germany}
\author{R.\ H\"artle}
\affiliation{Institut f\"ur theoretische Physik, Georg-August-Universit\"at G\"ottingen, Friedrich-Hund-Platz 1, D-37077 G\"ottingen, Germany}
\author{M.\ Thoss}
\affiliation{Institut f\"ur Theoretische Physik und Interdisziplin\"ares Zentrum f\"ur Molekulare Materialien, Friedrich-Alexander-Universit\"at Erlangen-N\"urnberg, Staudtstr.\ 7/B2, D-91058 Erlangen, Germany}

\date{\today}

\begin{abstract}
 	The interaction between electronic and vibrational degrees of freedom in single-molecule junctions may result from the dependence of the electronic energies or the electronic states of the molecular bridge on the nuclear displacement. The latter mechanism leads to a direct coupling between different electronic states and is referred to as nonadiabatic electronic-vibrational coupling. 
 	Employing a perturbative nonequilibrium Green's function approach, we study the influence of nonadiabatic electronic-vibrational coupling in model molecular junctions. Thereby we distinguish between systems with well separated and quasi-degenerate electronic levels.
	The results show that the nonadiabatic electronic-vibrational interaction can have a significant influence on the transport properties. The underlying mechanisms, in particular the difference between nonadiabatic and adiabatic electronic-vibrational coupling, are analyzed in some detail. 

\end{abstract}

\pacs{73.23.-b,85.65.+h,71.38.-k,72.10.Di}

\maketitle

\section{Introduction}

	Quantum transport in nanostructures is an active field of experimental and theoretical research. Among the variety of architectures investigated, single-molecule junctions, that is a single molecule chemically bound to two macroscopic leads, have been of great interest recently \cite{Reed97, Reichert02, Boehler2004, Venkataraman2006,  Ward08, Mohn10, Secker2010, Kim2011, Ballmann2013}. These systems combine the possibility to study fundamental aspects of nonequilibrium many-body quantum physics at the nanoscale with the perspective for applications in nanoelectronic devices \cite{Aviram1974, Nitzan_TopicalReview, Joachim2005, Cuniberti, Cuevas}. Studies of transport in molecular junctions have revealed a variety of interesting transport phenomena, such as rectification \cite{Braig2003, WuNazinHo04, Elbing05, Diez2009, Haertle11}, switching \cite{Blum05, Choi2006, Quek2009, Meded09, Benesch2009, Molen2010, Mohn10}, quantum interference \cite{Hod2006, Solomon2008_2, Begemann2008, Markussen2010,  Haertle11_3, Ballmann2012, Guedon2012, Haertle13} and negative differential resistances \cite{Gaudioso2000, Heinrich2001, Hettler2003, Koch2005, Pop2005, Sapmaz2006, Zazunov2006, Muralidhara2007, Tal2008, Leijnse2008, Haertle11}. 

	An important mechanism in electron transport across a molecular bridge is the coupling between the electronic and nuclear degrees of freedom \cite{Galperin_TopicalReview}. Due to the small size of molecules, transport-induced charge fluctuations influence the nuclear geometry. This leads to vibrational structures in the conductance of a molecular junction, current-induced vibrational excitation and a variety of interesting nonequilibrium effects \cite{Lauho2000, Mitra2004, Koch05, Galperin06, Sapmaz2006, Haertle10, Haertle11, Ballmann2013, Wilner14}. 
 	A well studied mechanism in this context is polaron-type transport, which results from the state-specific dependence of the electronic energies of the molecule on the nuclear displacement, that cause, e.g., a change of the potential energy surface of the molecular bridge upon charging of the molecule \cite{Galperin_TopicalReview,Haertle11}. 
	In addition, due to the dependence of the electronic states on the nuclear coordinates, the kinetic energy operator of the nuclei may cause transitions between different electronic states which can influence the transport properties profoundly. This mechanism, which represents a breakdown of the Born-Oppenheimer approximation, is referred to as nonadiabatic electronic-vibrational coupling \cite{Koeppel84,Domcke04} and has recently been investigated, e.g., in the context of Jahn-Teller effect in molecular junctions \cite{Reckermann2008, Frederiksen_JahnTeller_2008, Schultz2008, Han2010} and in STM studies of olygothiophene molecules on a Au substrate \cite{Repp2010}. 
	It also manifests itself in the off-resonant transport regime in structures in inelastic electron tunneling spectra (IETS) \cite{Frederiksen2005, Hihath2010, Avriller2012, Buerkle2013}. 
	In the present paper, we provide a model-based analysis of the mechanism of nonadiabatic electronic-vibrational coupling in molecular junctions transport and its interplay with polaron-type transport. 
	Thereby, we focus on the resonant transport regime, where the effects of electron-vibrational coupling are typically more pronounced. 

	A variety of different approaches have been used to describe vibrationally coupled electron transport in molecular junctions and other nanostructures, including density-matrix approaches \cite{May2002, Mitra2004, Pedersen2005, Harbola2006}, scattering theory \cite{Cizek2004, Caspary2007, Zimbovskaya2009, Jorn2009}, path integrals \cite{Muehlbacher2008, Weiss2008}, multiconfigurational wave-function methods \cite{Wang2009_2}, and nonequilibrium Green's functions theory \cite{Migdal1958, Caroli1972, Hyldgaard94, Pecchia2004, Galperin06,  Ryndyk2006, Galperin_TopicalReview, Ueda2007, Haertle08, Entin-Wohlman2009, Haupt2009}. In this work, we will apply the latter approach in combination with the self-consistent Born approximation to characterize transport in molecular junctions.
	
	The outline of the paper is as follows: In Sec.\ \ref{sec:theory} we introduce the model and the nonequilibrium Green's function approach used to describe vibrationally coupled transport through a molecular junction. 
	Sec.\ \ref{sec:results} presents results for selected model systems and an analysis of the basis effects of nonadiabatic electronic-vibrational coupling.
	Thereby, we distinguish between molecules with well separated (Sec.\ \ref{sec:normal}) and \mbox{(quasi-)}degenerate electronic states (Sec.\ \ref{sec:degenerate}). Sec.\ \ref{sec:conclusions} concludes with a summary.

\section{Theoretical methodology}
\label{sec:theory}

\subsection{Model Hamiltonian}\label{sec:hamiltonian}
	We consider charge transport through a molecule attached to two macroscopic leads. The molecule is described by the model Hamiltonian $H_{\text{Mol}}$, consisting of a set of discrete electronic and vibrational states which interact with each other.
	\begin{eqnarray}
	 \hspace*{-0.3cm}
	 H_{\text{Mol}} &=& \sum_\alpha \hbar\Omega_\alpha a_\alpha^\dagger a_\alpha + \sum_i \epsilon_i d_i^\dagger d_i + \sum_{\alpha i j} M_{ij}^\alpha Q_\alpha d_i^\dagger d_j \label{eq:Hamiltonian}
	\end{eqnarray}
	Thereby $\hbar\Omega_\alpha$ is the energy of the vibrational mode $\alpha$ described within the harmonic approximation and with the corresponding creation and annihilation operators, $a_\alpha^{\dagger}$, $a_\alpha$ and nuclear displacement 
	$Q_\alpha=(a_\alpha + a_\alpha^{\dagger})/\sqrt2$. Analogously, $\epsilon_i$ is the energy of the molecular electronic state $i$ with creation and annihilation operators $d_i^{\dagger}$, $d_i$. In our model, we neglect the spin of the electrons, electron-electron interactions as well as electronic-vibrational interactions of higher order in the nuclear displacement.

	Electron transport through a molecular junction influences the nuclear degrees of freedom of the molecule.
	Of special interest for the transport properties of a molecule is its response to charge fluctuations. Upon charging or decharging, the molecule will adapt its nuclear geometry, which can lead to excitation of the vibrational modes. On the other hand, a displacement of the nuclei affects the electronic structure of the molecule causing also a transition between different electronic states.
	These effects are accounted for by the last term in $H_{\text{Mol}}$, which describes the electronic-vibrational interaction with coupling strength $M_{ij}^\alpha$ and is assumed to be linear in the nuclear displacement $Q_\alpha$. 

	Often, the electronic-vibrational coupling is derived from the dependence of the electronic energies on the nuclear coordinates. This results in an electronic-vibrational coupling term that only depends on the population of the electronic states $d_i^\dagger d_i$ \cite{Galperin_TopicalReview}. This treatment corresponds to the adiabatic or Born-Oppenheimer approximation \cite{Cederbaum1974, Domcke04}. Without coupling to the leads, the resulting molecular Hamiltonian can be diagonalized using the polaron transformation \cite{Lang63, Mahan, Mitra04, Haertle08, Haertle11}. Within the adiabatic approximation, vibronic transitions between different electronic states of the molecule are only possible via coupling to the leads (charging and decharging) and the resulting transport phenomena can be rationalized within the Franck-Condon framework \cite{Koch2006, Haertle08, Haertle09, Romano2010, Haertle10, Haertle11, Haertle11_2, Volkovich2011, Haertle13}. 
	
	The nondiagonal electronic-vibrational interaction terms, $M_{ij}^\alpha Q_\alpha d_i^\dagger d_j$, $i \neq j$, on the other hand, describe transitions between different electronic states of the molecule induced by coupling to the nuclear degrees of freedom. These terms arise from the dependence of the electronic states on the nuclear degrees of freedom. They represent a breakdown of the adiabatic or Born-Oppenheimer approximation and are referred to as nonadiabatic coupling.  In  the context of nonadiabatic processes in molecules, these terms are often termed vibronic coupling \cite{Koeppel84, Domcke04}. Different electronic basis sets may be used to describe nonadiabatic coupling. While in the basis of the molecular eigenstates (so called adiabatic basis) nonadiabatic coupling results from the kinetic energy operator of the nuclei, the description used in this paper corresponds to the diabatic representation, where the nonadiabatic coupling is described by the nondiagonal elements of the diabatic potential matrix \cite{Domcke04}. Specifically, the molecular Hamiltonian in Eq.\ (\ref{eq:Hamiltonian}) corresponds to the well known vibronic coupling model and describes, e.g. conical intersections of potential energy surfaces and the associated phenomena \cite{Koeppel84, Domcke04}. 
	In Sec.\ \ref{sec:results} we identify differences between adiabatic and nonadiabatic coupling mechanisms effects in charge transport.

	The left and the right leads of the molecular junction are modeled as macroscopic reservoirs of noninteracting electrons,
	\begin{eqnarray}
	 H_{\text{L / R}} &=& \sum_{k \in \text{L/R}} \epsilon_k c_k^\dagger c_k,
	\end{eqnarray}
	where $\epsilon_k$ is the energy of leads state $k$ and $c_k^{\dagger}$,  $c_k$ denote the corresponding creation and annihilation operators. The coupling between the electronic states in the molecule and in the leads is described by a tight-binding like Hamiltonian
	\begin{eqnarray}
	 H_{\text{ML / MR}} &=& \sum_{k \in \text{L/R}, i} V_{ki} c_k^\dagger d_i + \text{h.c.},
	\end{eqnarray}
	with coupling constants $V_{ki}$.

	The Hamiltonian of the overall junction, comprising the molecule and the left and right leads, is given by
	\begin{eqnarray}
	 H &=& H_{\text{Mol}} + H_{\text{L}} + H_{\text{R}} + H_{\text{ML}} + H_{\text{MR}}.
	\end{eqnarray}

\subsection{Nonequilibrium Green's function theory}\label{sec:GF_theory}

	To describe the steady state transport properties of a molecular junction with electronic-vibrational coupling, we employ the nonequilibrium Green's function approach 
	based on perturbation theory introduced by Kadanoff, Baym, Keldysh and Langreth \cite{Kadanoff, Keldysh, Langreth, Mahan, Frederiksen, Galperin_TopicalReview, Haug/Jauho}. The central objects in describing the nonequilibrium many-body system are the electronic and vibrational Green's functions defined as:
	\begin{eqnarray}
		G_{ij}(\tau,\tau') = -\frac{\text{i}}{\hbar} \langle T_c d_i(\tau) d_j^\dagger(\tau') \rangle \\
		D_{\alpha\beta}(\tau,\tau') = -\frac{\text{i}}{\hbar} \langle T_c Q_\alpha(\tau) Q_\beta(\tau') \rangle
	\end{eqnarray}
	$G_{ij}$ denotes the electronic Green's functions and $i$, $j$ label molecular electronic states whereas $D_{\alpha\beta}$ stands for the vibrational Green's functions and $\alpha$, $\beta$ are vibrational modes. $T_c$ is the contour ordering operator in the complex time domain.
	The Green's functions as well as their retarded and advanced projections obey Dysons's equation which reads in energyspace
	\begin{eqnarray}
	 G_{ij}^{(r/a)}(\epsilon) = G_{ij}^{0 (r/a)}(\epsilon) + \sum_{kl} G_{ik}^{0 (r/a)}(\epsilon) \Sigma_{kl}^{(r/a)}(\epsilon) G_{lj}^{(r/a)}(\epsilon).\nonumber \\ \label{eq:Dyson}
	\end{eqnarray}
	The greater and lesser Green's function satisfy Keldysh's equation
	\begin{eqnarray}
	 G_{ij}^\lessgtr(\epsilon) &=& \sum_{kl}G_{ik}^{r}(\epsilon) \Sigma_{kl}^\lessgtr(\epsilon) G_{lj}^{a}(\epsilon).	\label{eq:Keldysh}
	\end{eqnarray}
	$G_{ij}^{0 (r/a)}(\epsilon)$ denotes the Green's function of the unperturbed system, without interactions or couplings. $k$ and $l$ label the molecular electronic states. $\Sigma_{ij}^{r/a/</>}(\epsilon)$ is the electronic self-energy accounting for the interactions and couplings. In this model, the self-energy consists of three parts describing the coupling to the left and right lead respectively and a part accounting for the electronic-vibrational interaction
	\begin{eqnarray}
	 \Sigma_{ij}(\epsilon) = \Sigma_{\text{L }ij}(\epsilon) + \Sigma_{\text{R }ij}(\epsilon) + \Sigma_{\text{vib }ij}(\epsilon) .
	\end{eqnarray}

	We describe the leads as semi-infinite tight-binding chains with inter-site coupling $\beta$ \cite{Cizek2004, PhDTutorial, Cuevas}. The self-energy for the coupling to the leads reads
	\begin{eqnarray}
	 \Sigma_{\text{L/R } ij}^{r/a} (\epsilon) &=& \Delta_{\text{L/R} ij}(\epsilon) \mp  \frac{\text{i}}{2} \Gamma_{\text{L/R} ij}(\epsilon),  \\
	 \Sigma_{\text{L/R } ij}^{<} (\epsilon) &=&  \text{i} \Gamma_{\text{L/R} ij}(\epsilon) f(\epsilon-\mu_\text{L/R}),  \\
	 \Sigma_{\text{L/R } ij}^{>} (\epsilon) &=&  -\text{i} \Gamma_{\text{L/R} ij}(\epsilon) (1-f(\epsilon-\mu_\text{L/R})).  
	\end{eqnarray}
	Here, $f(\epsilon)$ denotes the Fermi distribution function. The level-width function $\Gamma_{\text{L/R }ij}$  and the level-shift function $\Delta_{\text{L/R }ij}$ have the form
	\begin{eqnarray}
	 \Gamma_{\text{L/R} ij}(\epsilon) &=& \frac{V_{\text{L/R} i}V_{\text{L/R} j}^*}{\beta^2} \Theta(4\beta^2-x^2) \sqrt{4\beta^2-x^2} , \\
	 \Delta_{\text{L/R} ij}(\epsilon) &=& \frac{V_{\text{L/R} i}V_{\text{L/R} j}^*}{2\beta^2}\left( x -\Theta(x^2-4\beta^2) \sqrt{x^2 - 4\beta^2} \right),\nonumber \\ 
	\end{eqnarray}
	respectively,
	with $x = \epsilon-\mu_{\text{L/R }}$ and the Heaviside step function $\Theta$. We assume that an applied bias voltage leads to a symmetric change in the chemical potentials $\mu_\text{L/R}=\pm e\frac{V}{2}$.

	We use the standard self-consistent Born approximation (SCBA) to describe the electronic-vibrational interaction
	\cite{Migdal1958, Caroli1972, Mahan, Hyldgaard94, Frederiksen, Galperin_PeaksAndDips, Galperin_TopicalReview, Haug/Jauho}. 
	The corresponding self-energy is given by
	\begin{eqnarray}
	 \Sigma_{\text{vib }ij}^r(\epsilon) &=& \text{i} \sum_{a,b \in \text{mol} \atop \alpha \in \text{vib}} M_{ia}^{\alpha} M_{bj}^{\alpha} \int \frac{\text{d}\epsilon'}{2\pi} \Big( 
							  D_{\alpha\alpha}^{0r}(\epsilon-\epsilon') G_{ab}^r(\epsilon') \nonumber \\&&
							+ D_{\alpha\alpha}^{0<}(\epsilon-\epsilon') G_{ab}^r(\epsilon') 
							+ D_{\alpha\alpha}^{0r}(\epsilon-\epsilon') G_{ab}^<(\epsilon') \Big) \nonumber \\
					    & & -\text{i} \sum_{a,b \in \text{mol} \atop \alpha \in \text{vib}} M_{ij}^{\alpha} M_{ba}^{\alpha} \int \frac{\text{d}\epsilon'}{2\pi} \Big( 
							  D_{\alpha\alpha}^{0<}(\epsilon=0) G_{ab}^a(\epsilon') \nonumber \\&&
							+ D_{\alpha\alpha}^{0r}(\epsilon=0) G_{ab}^<(\epsilon') \Big), \label{eq:SE_SCBA:ret} \\  
	 \Sigma_{\text{vib } ij}^\lessgtr(\epsilon) &=& \text{i} \hspace{-0.25cm} \sum_{a,b \in \text{mol} \atop \alpha \in \text{vib}} \hspace{-0.25cm} M_{ia}^{\alpha} M_{bj}^{\alpha} \int \hspace{-0.1cm} \frac{\text{d}\epsilon'}{2\pi} \Big( 
							  D_{\alpha\alpha}^{0\lessgtr}(\epsilon-\epsilon') G_{ab}^\lessgtr(\epsilon')  \Big).  \label{eq:SE_SCBA:lessgtr}
	\end{eqnarray}	
	The self-energy consists of two parts, the Hartree and the Fock term. Unlike bulk solid state systems, the Hartree term, which is given by the terms proportional to $D_{\alpha\alpha}^{0r/<}(\epsilon=0)$ in Eq.\ (\ref{eq:SE_SCBA:ret}) and leads to renormalization of the electronic energies, does not vanish since the translational symmetry of the systems under consideration is broken \cite{Schrieffer, Hyldgaard94}. 
	We assume that the vibrations remain in thermal equilibrium throughout the transport process, which is indicated by $D_{\alpha\alpha}^{0r/\lessgtr}$. Approaches including nonequilibrium vibrational effects have been used, for example, in Refs.\ \onlinecite{Mitra04, Ryndyk2006, Galperin_PeaksAndDips}.
	Using Eqs.\ (\ref{eq:Dyson}),(\ref{eq:Keldysh}) and the definition of the self-energy in Eqs.\ (\ref{eq:SE_SCBA:ret}),(\ref{eq:SE_SCBA:lessgtr}), we obtain a closed set of equations which is solved iteratively.

	For characterizing the transport properties of a molecular junction, the main observable of interest is the electronic current. It can be calculated using the Meir-Wingreen-like formula \cite{Meir92} 
	\begin{eqnarray}
	 \hspace*{-0.3cm}
	 I_\text{L} &=& \frac{e}{2\pi\hbar} \hspace{-0.1cm} \sum_{ij} \hspace{-0.1cm} \int \hspace{-0.2cm} \Big( \Sigma_{\text{L }ij}^<(\epsilon) G_{ji}^>(\epsilon) - \Sigma_{\text{L }ij}^>(\epsilon) G_{ji}^<(\epsilon) \Big) \text{d}\epsilon.
	\end{eqnarray}
	Generally, the current $I_{\text{L}} = I_{\text{L el}} + I_{\text{L inel}}$ can be separated into an elastic part $I_{\text{L el}}$ and an inelastic part $I_{\text{L inel}}$ with
	\begin{eqnarray}
	 I_\text{L el} &=& \frac{e}{2\pi\hbar} \sum_{ijkl} \int \Big( \Sigma_{\text{L }ij}^<(\epsilon) G_{jk}^r(\epsilon)\Sigma_{\text{R }kl}^>(\epsilon)G_{li}^a(\epsilon) \nonumber \\ &&
			- \Sigma_{\text{L }ij}^>(\epsilon) G_{jk}^r(\epsilon)\Sigma_{\text{R }kl}^<(\epsilon)G_{li}^a(\epsilon) \Big) \text{d}\epsilon, 
	\end{eqnarray}
	\begin{eqnarray}
	 I_\text{L inel} &=& \frac{e}{2\pi\hbar} \sum_{ijkl} \int \Big( \Sigma_{\text{L }ij}^<(\epsilon) G_{jk}^r(\epsilon)\Sigma_{\text{vib }kl}^>(\epsilon)G_{li}^a(\epsilon)  \nonumber \\ &&
			 - \Sigma_{\text{L }ij}^>(\epsilon) G_{jk}^r(\epsilon)\Sigma_{\text{vib }kl}^<(\epsilon)G_{li}^a(\epsilon) \Big) \text{d}\epsilon.  \label{eq:inelastic_current}
	\end{eqnarray}
	Following \cite{Galperin_PeaksAndDips, Cuevas}, applying Dyson's Eq.\ (\ref{eq:Dyson}), the elastic current can be split into an electronic part and elastic correction part with
	\begin{eqnarray}
	 I_{\text{L el}}^0 &=& 
		\frac{e}{2\pi\hbar} \sum_{ijkl} \int \Big( \Sigma_{\text{L }ij}^<(\epsilon) G_{jk}^{0 r}(\epsilon)\Sigma_{\text{R }kl}^>(\epsilon)G_{li}^{0 a}(\epsilon)  \nonumber \\ &&
				- \Sigma_{\text{L }ij}^>(\epsilon) G_{jk}^{0 r}(\epsilon)\Sigma_{\text{R }kl}^<(\epsilon)G_{li}^{0 a}(\epsilon) \Big) \text{d}\epsilon, \\
	 \delta I_{\text{L el}} &=& I_{\text{L el}} - I_{\text{L el}}^0. \label{eq:elastic_corrections}
	\end{eqnarray}
	$I_{\text{L el}}^0$ is the electronic current, that is the current in absence of electronic-vibrational interactions. $\delta I_{\text{L el}}$ is the elastic correction introduced by the electronic-vibrational coupling. In lowest order, this contribution is proportional to $\Sigma_{\text{vib }ij}^{0 r/a}(\epsilon)$ and describes processes including the emission and reabsorption of virtual vibrational quanta \cite{Davis1970, Cuevas}.

\subsection{Nonequilibrium Green's function theory and interference effects}\label{sec:interference}

	The Green's function approach offers a description of the electronic transport including quantum coherences. Therefore it accounts for interference effects between different transport channels, which have received significant attention recently \cite{Haertle11_3, Ballmann2012, Vazquez2012}. 
	Thereby different approaches have been used for analysis, including the use of the eigenbasis of the molecular subspace \cite{Solomon2006, Haertle11_3} or transformations to conducting orbitals \cite{Solomon2008}. In this work, we apply the approach used in Ref.\ \onlinecite{Haertle13} to identify interference effects in transport, which employs the basis of molecular states to express the transmission function
	\begin{eqnarray}
	 T(\epsilon) = \sum_{ij} \Gamma_{L ij}(\epsilon) G_{ji}^>(\epsilon).
	\end{eqnarray}
	Within this framework, the incoherent contribution to transport is obtained by neglecting the offdiagonal parts of the self-energy of the coupling to the left lead,
	\begin{eqnarray}
	 T_{\text{incoh}}(\epsilon) = \sum_{i} \Gamma_{L ii}(\epsilon) G_{ii}^>(\epsilon). \label{eq:transmission_incoherent}
	\end{eqnarray}
	This corresponds to a coupling of the electronic states to separate electrodes, excluding the effect of interference.
	Accordingly, the part of the transmission function describing interference effects reads
	\begin{eqnarray}
	 T_{\text{interf}}(\epsilon) = T(\epsilon) - T_{\text{incoh}}(\epsilon) = \sum_{i, j\neq i} \Gamma_{L ij}(\epsilon) G_{ji}^>(\epsilon). \label{eq:transmission_interference}
	\end{eqnarray}

\section{Results}\label{sec:results}

		\begin{table*}[tb]
		 \caption{Parameters describing the model systems investigated in this article. For all calculations, the temperature is $T=10$K. All parameters are given in eV.}
 	  	\begin{center}
		\begin{tabular}{l||c|c|c|c|c|c|c|c|c|c|c}
		Model  & $\epsilon_1$ & $\epsilon_2$ & $V_{\text{L }1}$ & $V_{\text{R }1}$ & $V_{\text{L }2}$ & $V_{\text{R }2}$ & $\beta$ & $\hbar\Omega$ & $M_{11}$ & $M_{22}$ & $M_{12}$ \\ \hline\hline
		AD    & $0.25$ & $0.4$ & $0.1$ & $0.1$ & $0.1$ & $0.1$ & $3$ & $0.025$--$0.3$ & $0.03$ & $0.03$ & $0.0$ \\
		NONAD & $0.25$ & $0.4$ & $0.1$ & $0.1$ & $0.1$ & $0.1$ & $3$ & $0.025$--$0.3$ & $0.0$ & $0.0$ & $0.03$ \\
		ASYMM & $0.25$ & $0.4$ & $0.02$ & $0.2$ & $0.2$ & $0.02$ & $3$ & $0.1$ & $0.0$ & $0.0$ & $0.03$ \\
		INTPLY & $0.25$ & $0.4$ & $0.1$ & $0.1$ & $0.1$ & $0.1$ & $3$ & $0.1$ & $0.05$ & $0.05$ & $0.0$, $\pm 0.02$ \\
		DESNONAD& $0.5$ & $0.3$ -- $0.7$ & $0.1$ & $0.1$ & $0.1$ & $-0.1$ & $2$ & $0.1$ & -- & -- & $0$, $0.05$ \\
		DESVIB & $0.5$ & $0.505$ & $0.1$ & $0.1$ & $0.1$ & $-0.1$ & $2$ & $0.1$ & $0.0$ & $0.05$ & $0.0$ -- $0.02$ \\
		\end{tabular}
 		\end{center}
		\label{tab:parameters}
		\end{table*}
	In this section, we analyze the influence of nonadiabatic electronic-vibrational coupling on the transport properties of model systems. The parameters characterizing the model systems, summarized in Tab.\ \ref{tab:parameters}, represent typical values for molecular junctions. We emphasize, however, that we restrict ourselves to parameter regimes where perturbation theory is valid.

	This section is separated into two parts, depending on the energy spacing between the electronic levels. In part \ref{sec:normal} we discuss the effect of electronic-vibrational coupling for two-level systems with well separated electronic states including an analytic study of important vibrational effects in Sec.\ \ref{sec:normal:identification}, a comparison of the influence of adiabatic and nonadiabatic vibrational interactions in Sec.\ \ref{sec:normal:ad_vs_nonad} and an investigation of the transport properties of model systems with increased complexity in Sec.\ \ref{sec:normal:ad_vs_nonad} -- \ref{sec:normal:interpaly}.
	In the second part \ref{sec:degenerate}, we study the influence of electronic-vibrational interactions in two-level systems with quasi-degenerate electronic levels. For such systems, quantum interference effects are important and sensitive to the electronic-vibrational interaction \cite{Hod2006, Begemann2008, Haertle11_3, Ballmann2012, Haertle13}.

\subsection{Transport properties of molecular conductors with well separated electronic states}\label{sec:normal}

	In this section we investigate the effect of nonadiabatic electronic-vibrational coupling on the transport properties of molecular junctions with well separated electronic levels, i.e. with energy spacings that exceed their broadening induced by the coupling to the leads. For the sake of clarity, we restrict our discussion to systems with a single vibrational mode, dropping the vibrational indices. A generalization to multimode systems is, in principle, straightforward.

\subsubsection{Identification of dominant processes}\label{sec:normal:identification}

	We first analyze the current, as the main transport observable. Since the method outlined in Sec.\ \ref{sec:GF_theory} restricts the investigation to model systems where perturbation theory is valid, we perform an expansion of the current in $M_{ij}$ and $V_{ij}$ to lowest nonvanishing order. This allows us to systematically identify the most important transport processes. 
 
	To this end, we introduce the shorthand notations
	\begin{eqnarray}
	 \mathcal{G}_{ij}^r(\epsilon) &=& \text{i} \int \frac{\text{d}\epsilon'}{2\pi} \Big( 
							  D^{0r}(\epsilon-\epsilon') G_{ij}^r(\epsilon')  
							+ D^{0<}(\epsilon-\epsilon') G_{ij}^r(\epsilon') \nonumber \\ && 
							+ D^{0r}(\epsilon-\epsilon') G_{ij}^<(\epsilon') \Big), \label{eq:Shorthand_r}
	\end{eqnarray}
	\begin{eqnarray}
	 \mathcal{G}_{ij}^\lessgtr(\epsilon) &=& 
						\text{i} \sum_{kl}\int \frac{\text{d}\epsilon'}{2\pi} \Big( D^{0\lessgtr}(\epsilon-\epsilon') G_{ik}^r(\epsilon') \Sigma_{kl}^\lessgtr(\epsilon') G_{lj}^a(\epsilon') \Big), \nonumber \\ \label{eq:Shorthand_lessgtr}
	\end{eqnarray}
	which appear in the definition of the vibrational self-energies Eq.\ (\ref{eq:SE_SCBA:ret}) and (\ref{eq:SE_SCBA:lessgtr}). A lowest order estimate of $\mathcal{G}_{ij}^{r \lessgtr}(\epsilon)$ can be obtained by inserting the expressions for free particle Green's functions into the right hand side of Eq.\ (\ref{eq:Shorthand_r}) and (\ref{eq:Shorthand_lessgtr}). We identify $\mathcal{G}_{ij}^{0r}(\epsilon) \hat{=} G_{ij}^{0r}(\epsilon-\hbar\Omega)$ and $\mathcal{G}_{ij}^{0 \lessgtr}(\epsilon) \hat{=} G_{ij}^{0 \lessgtr}(\epsilon-\hbar\Omega)$ in this order.

	Vibrational effects in the current appear in the elastic corrections (\ref{eq:elastic_corrections}) and the inelastic current (\ref{eq:inelastic_current}). For weak coupling, the most important vibrational effects are of lowest nonvanishing order in the electronic-vibrational and the molecule-lead coupling, $\mathcal{O}(M^2V^4)$. 
	To this order and for low temperature, the parts of the current influenced by the vibrations read
	\begin{widetext}
	\begin{eqnarray}
 	 \delta I_{\text{L el}} &=& \frac{e}{2\pi\hbar} \sum_{ijkab} \int \Big( 
				\Sigma_{\text{L }ij}^<(\epsilon) \left( G_{jj}^{0r}(\epsilon) M_{ja} M_{bk} \mathcal{G}_{ab}^{0r}(\epsilon) G_{kk}^{0r}(\epsilon) G_{ii}^{0a}(\epsilon) + \text{h.c.} \right) \Sigma_{\text{R }ki}^>(\epsilon) 
			\Big) \text{d}\epsilon \nonumber \\  
				& & -\frac{e}{2\pi\hbar} \sum_{ijka} \int \Big( 
				\Sigma_{\text{L }ij}^<(\epsilon) \left( G_{jj}^{0r}(\epsilon) M_{ja} M_{ak} \frac{2n_{aa}}{\hbar\Omega} G_{kk}^{0r}(\epsilon)  G_{ii}^{0a}(\epsilon) + \text{h.c.} \right) \Sigma_{\text{R }ki}^>(\epsilon) \Big) \text{d}\epsilon + \mathcal{O}(M^4 V^6), \label{eq:normal:identification1}\\
	 I_{\text{L inel}} &=& \frac{e}{2\pi\hbar} \sum_{ijab} \int \Big( 
			\Sigma_{\text{L }ji}^<(\epsilon) G_{ii}^{0 r}(\epsilon) M_{ia} M_{bj} \mathcal{G}_{ab}^{0>}(\epsilon) G_{jj}^{0 a}(\epsilon)
		\Big) \text{d}\epsilon + \mathcal{O}(M^4 V^6), \label{eq:normal:identification2}
	\end{eqnarray}
	\end{widetext}
	where $n_{aa}$ denotes the population of the electronic state $a$.
	This expression can be further simplified for well separated electronic levels. Under this assumption, the product of two Green's functions at the same energy but for different electronic states is negligible. Note that this approximation does not rely on perturbation theory but uses the fact that the electronic Green's functions are strongly peaked at their resonances. Applying this simplification, Eq.\ (\ref{eq:normal:identification1}) and (\ref{eq:normal:identification2}) reduce to
 	\begin{widetext}
	\begin{eqnarray}
 	 \delta I_{\text{L el}} &\approx& \frac{e}{2\pi\hbar} \sum_{i} \Bigg( \int 
				\Sigma_{\text{L }ii}^<(\epsilon) \Big( |G_{ii}^{0r}(\epsilon)|^2 G_{ii}^{0r}(\epsilon) \big[ M_{ii}^2 \mathcal{G}_{ii}^{0r}(\epsilon) + \sum_{a\neq i} |M_{ia}|^2 \mathcal{G}_{aa}^{0r}(\epsilon) \big] + \text{h.c.} \Big) \Sigma_{\text{R }ii}^>(\epsilon) 
			\nonumber \\
				& & -
				\Sigma_{\text{L }ii}^<(\epsilon) \Big( |G_{ii}^{0r}(\epsilon)|^2 G_{ii}^{0r}(\epsilon) \frac{2}{\hbar\Omega} \big[ M_{ii}^2 n_{ii} + \sum_{a\neq i} |M_{ia}|^2 n_{aa} \big] + \text{h.c.} \Big) \Sigma_{\text{R }ii}^>(\epsilon)
			\text{d}\epsilon \Bigg)
 		+ \mathcal{O}(M^4 V^6),  \label{eq:identification:el_corr} \\
	 I_{\text{L inel}} &\approx& \frac{e}{2\pi\hbar} \sum_{i} \int 
			\Sigma_{\text{L }ii}^<(\epsilon) |G_{ii}^{0 r}(\epsilon)|^2 \big[ M_{ii}^2 \mathcal{G}_{ii}^{0>}(\epsilon) + \sum_{a\neq i} |M_{ia}|^2 \mathcal{G}_{aa}^{0>}(\epsilon) \big] 
		\text{d}\epsilon + \mathcal{O}(M^4 V^6). \label{eq:identification:inel}
	\end{eqnarray}
	\end{widetext}

	The influence of the vibrations on the current separates into two parts. The first one, proportional to $M_{ii}^2 \mathcal{G}_{ii}^{0r/>}(\epsilon)$, depends only on the adiabatic electronic-vibrational coupling.
	The second part is proportional to $|M_{ia}|^2 \mathcal{G}_{aa}^{0r/>}(\epsilon)$ with $a\neq i$ and depends only on the nonadiabatic coupling strengths. Although the analytic description of these terms is analogous to the purely adiabatic terms, the nonadiabatic processes depend on two different electronic levels leading to a distinct dependence on the energies $\epsilon_i$, $\epsilon_a$ and $\hbar\Omega$, which will be exemplified in the following sections. Pronounced nonadiabatic vibrational effects appear if the resonance condition $\epsilon_i \approx \epsilon_a + \hbar\Omega$ is satisfied.
	This behavior is qualitatively different from the Franck-Condon picture for adiabatic electronic-vibrational effects. Accordingly, the adiabatic approximations is valid as long as the nonadiabatic coupling strengths $M_{ia}$ are small and the resonance condition is not fullfilled for any pair of the electronic states $i$ and $a$. This is reminiscent of the validity of the Born-Oppenheimer approximation. Notice that beyond the lowest order approximation in the molecule-lead coupling, the exact form of the level broadening provided by the leads influences the impact of the individual vibrational effects.
	Since in Eq.\ (\ref{eq:identification:el_corr}) and (\ref{eq:identification:inel}) the electronic-vibrational coupling decomposes in a purely adiabatic and a purely nonadiabatic part, we start the model based investigation of the nonadiabatic electronic-vibrational coupling below by a comparison between the well known adiabatic effects and the influence of purely nonadiabatic electronic-vibrational couplings in Sec.\ \ref{sec:normal:ad_vs_nonad} and \ref{sec:normal:new_channels}.
	
	Neglecting the product between two Green's functions at the same energy but for different electronic states is an approximation, which does not hold, for example, in systems where the coupling to the leads allows for strong mixing between the different electronic states, resulting in large off-diagonal components of the self-energy for the leads. In Appendix \ref{appendix:normal:identification}, we go beyond this approximation and analyze higher order processes
	as well as the interplay between adiabatic and nonadiabatic electronic-vibrational interactions. The latter is of importance for Sec.\ \ref{sec:normal:interpaly}.

\subsubsection{Comparison between adiabatic and nonadiabatic electronic-vibrational coupling}\label{sec:normal:ad_vs_nonad}

	To analyze the mechanisms and signatures of adiabatic and nonadiabatic electronic-vibrational coupling, we first consider the two effects separately. To this end, we compare results for two models subject to nonadiabatic electronic-vibrational coupling (labeled NONAD) and adiabatic electronic-vibrational coupling (AD), respectively. The parameters defining these model systems are given in Tab.\ \ref{tab:parameters}. 

\begin{figure}[tb]
		\hspace{-0.5cm}
 		\includegraphics[width=0.5\textwidth]{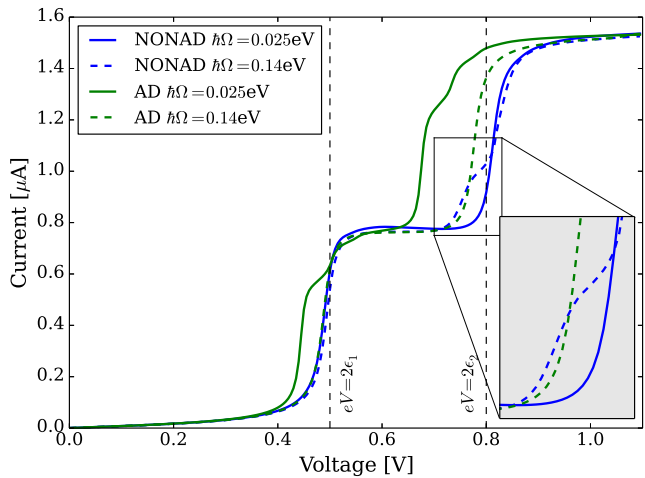}
		\caption{Current-voltage characteristics of two molecular junctions described by the models AD and NONAD for two different vibrational energies. 
		The onset of resonant transport through the electronic levels in the absence of vibrations is marked by vertical dashed lines. The inset provides a closeup on a region relevant for nonadiabatic vibrational effects.}
		\label{fig:normal:ad_vs_nonad:IV_char}
	\end{figure}

\begin{figure}[tb]
		\hspace{-0.5cm}
 		\includegraphics[width=0.5\textwidth]{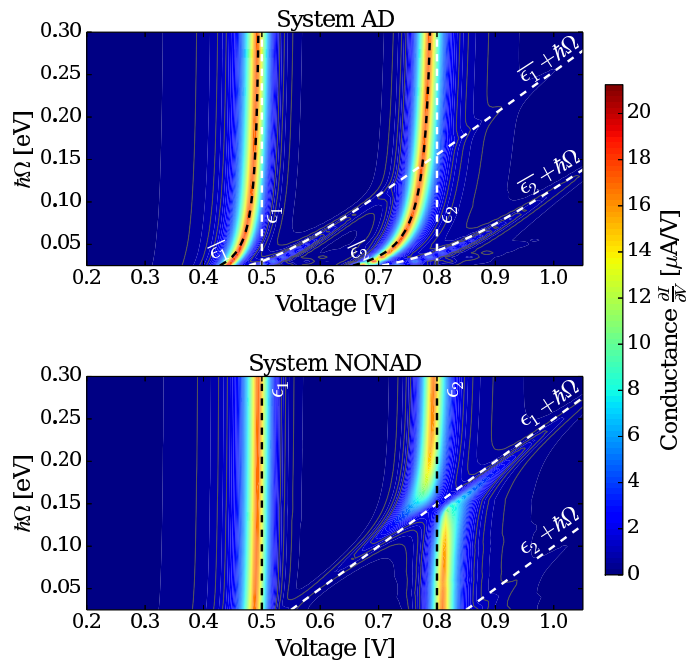}
		\caption{Conductance map for the model systems AD (top) and NONAD (bottom). In both plots, the conductance $\frac{\partial I}{\partial V}$ is plotted as a function of voltage $V$ and and vibrational energy $\hbar\Omega$. The dashed lines indicate the positions of the unperturbed electronic resonances $\epsilon_{1/2}$, the electronic resonances renormalized by the coupling to the vibrations $\overline{\epsilon}_{1/2}$ as well as their respective vibrational satellites.}
		\label{fig:normal:ad_vs_nonad:conductance_map}
	\end{figure}

	Fig.\ \ref{fig:normal:ad_vs_nonad:IV_char} shows the current-voltage characteristics for the model systems AD and NONAD for two different vibrational energies $\hbar\Omega = 0.025$eV and $\hbar\Omega = 0.14$eV.  The vibrational energies are chosen such that significant adiabatic ($\hbar\Omega = 0.025$eV) and nonadiabatic vibrational effects ($\hbar\Omega = 0.14$eV) can be observed. 
	A more comprehensive comparison between adiabatic and nonadiabatic electronic-vibrational effects is provided by the conductance map in Fig.\ \ref{fig:normal:ad_vs_nonad:conductance_map}, which depicts the conductance $\frac{\partial I}{\partial V}$ as a function of voltage and vibrational energy $\hbar\Omega$.
	\begin{figure*}[tb]
		\centering
 		\includegraphics[width=\textwidth]{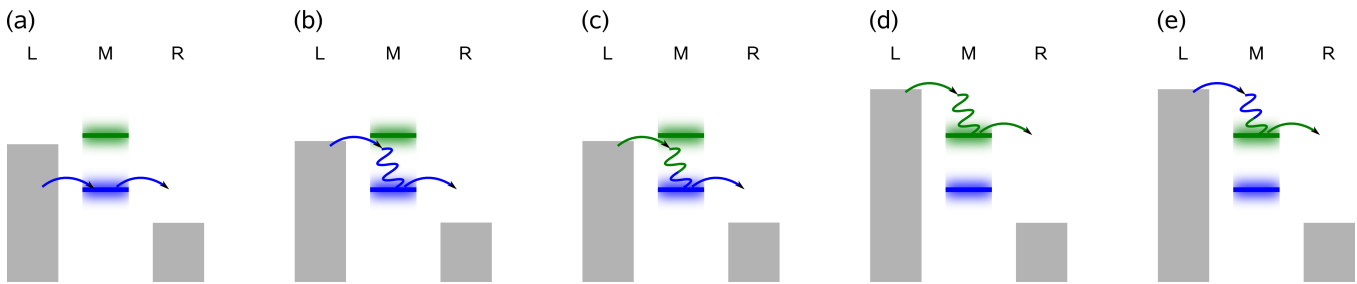}
		\caption{Schematic representation of example processes for purely electronic (a) and vibronic transport involving the emission of one vibrational quantum (b)--(e) and two broadened electronic levels.
 		The abbreviations L and R denote the left and the right lead respectively, and M stands for the molecule. 
		The arrows illustrate tunneling electrons whereas the color of the arrow denotes the molecular level responsible for the tunneling process.}
		\label{fig:normal:ad_vs_nonad:processes}
	\end{figure*}
	
	For both models, the current-voltage characteristics are dominated by two steps at different voltages which lead to lines in the conductance maps. 
	These features correspond to the onset of resonant transport through the electronic states as depicted in Fig.\ \ref{fig:normal:ad_vs_nonad:processes}(a) for the lower electronic state. 
	The position of the steps is determined by the energies of the electronic states renormalized by the coupling to the vibrations $\overline{\epsilon}_{1/2}$.
	In the purely adiabatic case, this renormalization is described by the polaron shift \cite{Lang63, Mahan, Haertle11} (black dashed lines in the conductance map of AD). 
	The electronic resonances of model NONAD, on the other hand, coincide very well with the unperturbed energies $\epsilon_{1/2}$ (black dashed lines in the conductance map of NONAD). As the purely nonadiabatic electronic-vibrational interaction provides a coupling between the electronic states, the energy shift due to the vibration depends on the energy difference between the electronic states (for details see Sec.\ \ref{sec:degenerate:constant_coupling}). The levels in NONAD are well separated, therefore the electronic energies are almost unchanged.

	Apart from the electronic resonances, the current exhibits additional smaller steps and the conductance additional, less pronounced lines at $eV = 2(\overline{\epsilon}_{1/2}+n\hbar\Omega)$ ($n\in \mathbf{N}$)(white dashed lines in the conductance maps). 
	Notice that vibrational features are clearly visible for $n=1$ only, since the SCBA is restricted to weak electronic-vibrational coupling.
	For the system AD, these features are more pronounced for small $\hbar\Omega$ (which results in a stronger dimensionless coupling $\frac{M_{ii}}{\hbar\Omega}$), whereas for the system NONAD they are only present for vibrational energies close to the resonance condition $\epsilon_1+\hbar\Omega = \epsilon_2$. In this region, the conductance map of NONAD also exhibits an avoided level crossing (intersection between white dashed line and black dashed line in Fig.\ \ref{fig:normal:ad_vs_nonad:conductance_map} (bottom)). These features correspond to the onset of transport processes, where electrons populate the molecular bridge resonantly upon the excitation of $n$ vibrational quanta. Examples for such processes are shown in Fig.\ \ref{fig:normal:ad_vs_nonad:processes}(b)--(e).

	In case of purely adiabatic electronic-vibrational coupling, these effects have been investigated in detail previously \cite{Mahan, Mitra04, Koch2006, Galperin_TopicalReview, Haertle08, Haertle09, Romano2010, Haertle10, Haertle11, Haertle11_2, Volkovich2011, Haertle13}.
	It was shown that the step height corresponding to vibronic processes can be described by the Franck-Condon factors and depends only on the dimensionless coupling $M_{ii}/\hbar\Omega$. As the processes only include one electronic level, the broadening of the electronic level and its proximity to its vibrational satellites is important for pronounced adiabatic vibrational effects in the current (see color coding in Fig.\ \ref{fig:normal:ad_vs_nonad:processes}(b) and (d)).
 
	Nonadiabatic electronic-vibrational processes, on the other hand, include two different electronic states. 
	In this process, depicted in Fig.\ \ref{fig:normal:ad_vs_nonad:processes}(c), one electronic state is populated from the left lead, whereas the other state allows the electron to leave the molecule towards the right lead. The intramolecular transition between the electronic states is facilitated by the coupling to the vibrations and the energy of the electron is changed by $\hbar\Omega$. The probability for this nonadiabatic process is enhanced if the resonance condition $\epsilon_1+\hbar\Omega = \epsilon_2$ is fulfilled \cite{Repp2010}.
	Because system NONAD comprises only a single vibrational degree of freedom, the nonadiabatic transport process results in an avoided level crossing as a signature of the breakdown of the Born-Oppenheimer approximation. In systems with more vibrational degrees of freedom, also true degeneracies are possible, for example in systems with conical intersection of potential energy surfaces \cite{Domcke04}.

\subsubsection{Transport channels due to nonadiabatic electronic-vibrational coupling and asymmetric current-voltage characteristics}\label{sec:normal:new_channels}

	As shown, nonadiabatic electronic-vibrational coupling provides a mechanism that mixes the electronic levels of the molecular bridge even without coupling to the leads. This induces additional transport channels, which are not available within the adiabatic approximation. A particular interesting scenario arises in junctions with asymmetric, state-specific coupling to the leads. As an example, we consider model ASYMM, where the lower electronic state couples strongly to the right lead and weakly to the left, whereas the other electronic level is strongly coupled to the left lead and weakly to the right. The corresponding parameters can be found in Tab.\ \ref{tab:parameters}.
	A system with similar coupling to the leads but without nonadiabatic electronic-vibrational coupling was investigated in \cite{Haertle10}.
	\begin{figure}[tb]
		\hspace{-0.5cm}
 		\includegraphics[width=0.5\textwidth]{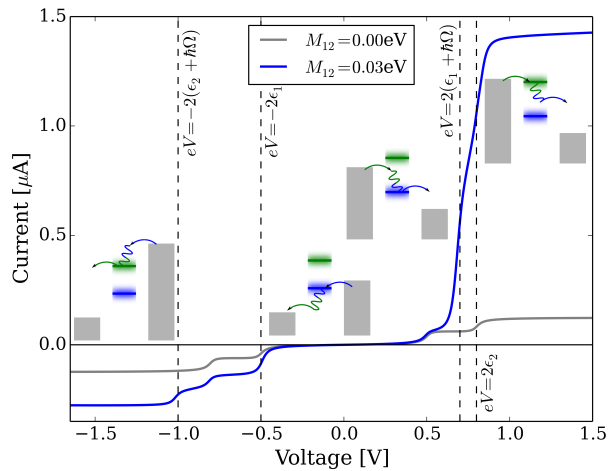}
		\caption{Current-voltage characteristics for the system ASYMM. 
			The pictures given in this plot visualize the most important vibrational transport processes. 
			The vertical dashed lines mark the voltages at which relevant unperturbed electronic energy levels enter the voltage window.}
		\label{fig:new_transport_channels}	
	\end{figure}

	Fig.\ \ref{fig:new_transport_channels} depicts the current-voltage characteristic of model ASYMM with and without electronic-vibrational coupling. 
	While the noninteracting model shows only a small current, the model with finite electronic-vibrational coupling, $M_{12}=0.03$eV, is a far better conductor with a pronounced inelastic current. 
	As is well known, the current-voltage characteristics for the  noninteracting model is, despite the asymmetry of the molecule lead coupling, symmetric with respect to bias polarity \cite{Galperin2005, Galperin_TopicalReview}. The nonadiabatic vibrational coupling, however, results in a pronounced asymmetry of the current-voltage characteristics. 
	The inelastic currents obtained for positive and negative bias polarity, respectively, do not only differ in magnitude, but also with respect to the voltages at which steps in the current appear.

	This is due to the fact that nonadiabatic electronic-vibrational interaction provides additional inelastic transport channels described by the expressions
	\begin{eqnarray}
	 I_{ij} &=& 	\frac{e}{2\pi\hbar} \int \Sigma_{\text{L }ii}^<(\epsilon) |G_{ii}^r(\epsilon)|^2 |M_{ij}|^2 \mathcal{G}_{jj}^>(\epsilon)\text{d}\epsilon \nonumber \\
	   &=&	\frac{e}{2\pi\hbar} \int \Sigma_{\text{L }ii}^<(\epsilon) |G_{ii}^{0r}(\epsilon)|^2 |M_{ij}|^2 |G_{jj}^{0r}(\epsilon-\hbar\Omega)|^2 \times \nonumber \\ && \times  \Sigma_{\text{R }jj}^>(\epsilon-\hbar\Omega) \text{d}\epsilon + \mathcal{O}(M^4 V^6).
	\end{eqnarray}
	for the current with $i,j \in \{1,2\}$, $i \neq j$. The additional steps in the inelastic current correspond to resonant transport processes depicted in Fig.\ \ref{fig:new_transport_channels}. 
 
	As the molecule couples asymmetrically to the leads, transport is dominated by inelastic channels that can exploit the strong couplings of the higher molecular electronic level to the left lead and of the lower molecular electronic level to the right lead. The transition between the electronic states is enabled by the interaction with the vibrations, where electrons scatter from the higher to the lower electronic level, thereby exciting the vibrational mode. Due to the low temperature of $T=10$ K, the reverse process including the absorption of vibrational energy is strongly suppressed. 
	For positive bias voltage, the asymmetric coupling to the leads together with the energetic restriction imposed by the emission of vibrational quanta results in a pronounced inelastic current that profits from the energetic proximity of $\epsilon_1 + \hbar\Omega$ and $\epsilon_2$. 
	For negative bias, on the other hand, the necessity to emit vibrational energy restricts the system to different inelastic transport processes, which do not benefit from the resonance condition found in Sec.\ \ref{sec:normal:identification}. This leads to a smaller current which exhibits steps at different voltages.

	This example shows that nonadiabatic electronic-vibrational coupling in junctions with asymmetric, state-specific molecule-lead coupling may results in a pronounced asymmetry of the current, i.e.\ a rectification effect. 
	It is noted that vibrationally induced rectification is also observed in similar models with adiabatic electronic-vibrational coupling \cite{Haertle11,Volkovich2011}, the underlying mechanism, however, is quite different.

\subsubsection{Interplay between adiabatic and nonadiabatic electronic-vibrational coupling}\label{sec:normal:interpaly}

	In real world molecules, both the adiabatic and nonadiabatic electronic-vibrational interactions will be active and will influence the transport properties in a more complex way than in the simplified models discussed above. The dominating contributions to the current from the adiabatic and nonadiabatic coupling mechanisms and their interplay in this more general scenario are analyzed in the Appendix. As an illustrative example, we consider model INTPLY, with parameters as specified in Tab.\ \ref{tab:parameters}.
	
	\begin{figure}[tb]
		\hspace{-0.5cm}
		\includegraphics[width=0.5\textwidth]{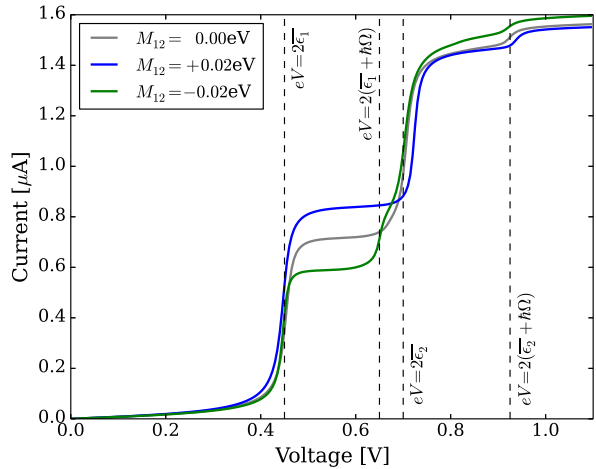}
		\caption{Current-voltage characteristics for the model system INTPLY for $M_{12}=0, \pm0.02$eV. 
			The vertical dashed lines denote the voltages at which resonant transport through the molecular states, renormalized by the adiabatic electronic-vibrational coupling, becomes possible.}
		\label{fig:normal:interplay}
	\end{figure}

	Fig.\ \ref{fig:normal:interplay} depicts the current-voltage characteristics of the model INTPLY for three different nonadiabatic electronic-vibrational coupling strengths $M_{12}$. The results reveal a significant influence of the nonadiabatic coupling in the voltage range between $eV = 2\overline{\epsilon}_1$ and $eV = 2(\overline{\epsilon}_1+\hbar\Omega)$, where $\overline{\epsilon}_1={\epsilon}_1-\frac{M_{11}^2}{\hbar\Omega}$ denotes the energy of the first electronic state renormalized by the adiabatic coupling to the vibrations (polaron shift). In particular, the current depends on the sign of the nonadiabatic coupling $M_{12}$.
	
\begin{figure}[tb]
		\hspace{-0.5cm}
		\includegraphics[width=0.5\textwidth]{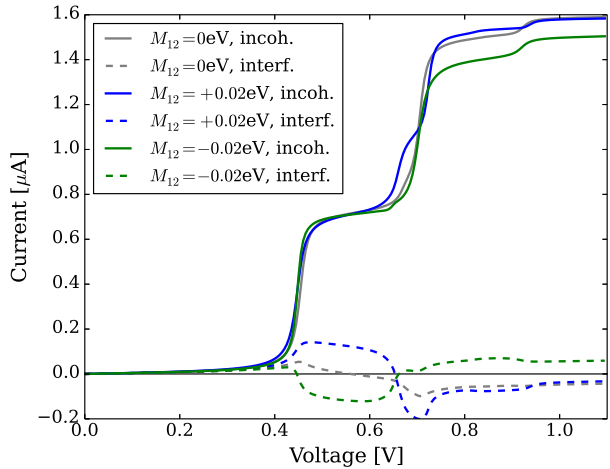}
		\caption{Interference (dashed lines) and incoherent (solid lines) contributions to the currents of system INTPLY for $M_{12}=0, \pm0.02$eV depicted in Fig.\ \ref{fig:normal:interplay}.}
		\label{fig:normal:interplay:identification}
	\end{figure}

	To analyze the underlying mechanism, in Fig.\ \ref{fig:normal:interplay:identification} we separate the current into an incoherent and an interference part, according to Eqs.\ (\ref{eq:transmission_incoherent}) and (\ref{eq:transmission_interference}). The results show that the change of the current induced by nonadiabatic coupling can be considered as an interference effect, whereby positive (negative) coupling  $M_{12}$ results in constructive (destructive) interference.
	Because the electronic levels of system INTPLY are well separated, the effect occurring in the voltage range between $eV = 2\overline{\epsilon}_1$ and $eV = 2(\overline{\epsilon}_1+\hbar\Omega)$ cannot be caused by interference of pathways associated with different electronic states. As is analyzed in more detail in the Appendix, it rather corresponds to interference of different vibronic contributions to the current induced by 	adiabatic and nonadiabatic electronic-vibrational coupling (cf.\ Eqs.\ (\ref{eq:appendix:1}) and (\ref{eq:appendix:2})). Thus, currents from adiabatic and nonadiabatic electronic-vibrational processes interfere with each other.
	For higher voltages, when the second electronic state enters the bias window, the effect disappears. This is because the self-energy for the coupling to the leads gives off-diagonal Green's function elements that differ in sign for the states $\epsilon_{1/2}$. 
 	As a result, transport processes where the role of the electronic states is interchanged contribute with opposite signs to the overall current and thus cancel each other.

	It is also noted that the interplay between adiabatic and nonadiabatic electronic-vibrational coupling breaks the symmetry of the current with respect to the first and second electronic state: Without nonadiabatic coupling, the transport through the two electronic levels contribute approximately equally to the current (for a system with the same molecule-lead and vibronic couplings for the two states as considered here and neglecting nonequilibrium vibrational effects and electron-electron interactions). 
	For a system with additional nonadiabatic coupling, the increase in the current at the bias voltages $eV = 2\overline{\epsilon}_1$ and $eV = 2\overline{\epsilon}_2$ differs
	due the interference between nonadiabatic and adiabatic contributions as discussed above.

\subsection{Transport properties of molecular junctions with \mbox{(quasi-)}degenerate electronic states}\label{sec:degenerate}
	In this section we investigate the effect of nonadiabatic electronic-vibrational coupling on model systems with \mbox{(quasi-)}degenerate electronic levels. Thereby we refer to molecular states as \mbox{(quasi-)}degenerate whenever the energy difference between the states is smaller than the level broadening due to coupling to the leads.
	For such a model system, quantum interference plays an important role and adiabatic electronic-vibrational coupling has been shown to result in decoherence effects \cite{Hod2006, Begemann2008, Haertle11_3, Ballmann2012, Haertle13}.

\subsubsection{Electronic two-level system with constant interstate coupling}\label{sec:degenerate:constant_coupling}

	The identification of dominant transport processes for systems with \mbox{(quasi-)}degenerate electronic states in analogy to the approach in Sec.\ \ref{sec:normal:identification} is significantly more involved than for well separated energy levels. To analyze their transport properties, we therefore consider a simplified toy model consisting of two electronic levels with constant coupling between the two electronic states at the molecular bridge instead of the vibronic coupling described by Eq.\ (\ref{eq:Hamiltonian}).  
	\begin{eqnarray}
	 H' &=& \epsilon_1 d_1^\dagger d_1 + \epsilon_2 d_2^\dagger d_2 + \Delta (d_1^\dagger d_2 + d_2^\dagger d_1) + \sum_{k \in \text{L/R}} \epsilon_k c_k^\dagger c_k \nonumber \\ &&
		+ \sum_{k \in \text{L/R}} \left( V_{k1} c_k^\dagger d_1 +  V_{k2} c_k^\dagger d_2 + \text{h.c.} \right) ,  \label{eq:degenerate:simple_Hamiltonian}
	\end{eqnarray}
	where $\Delta$ is the interstate constant coupling strength.
	The part of the Hamiltonian without the coupling to the leads can be diagonalized by a simple basis transformation, resulting in
	\begin{eqnarray}
	 	H' &=& \tilde{\epsilon}_1 \tilde d_1^\dagger \tilde d_1 + \tilde{\epsilon}_2 \tilde d_2^\dagger \tilde d_2 + \sum_{k \in \text{L/R}} \epsilon_k c_k^\dagger c_k \nonumber \\ &&
		+ \sum_{k\in \text{L/R}} \left( \tilde V_{k1} c_k^\dagger \tilde d_1 + \tilde V_{k2} c_k^\dagger \tilde d_2 + \text{h.c.} \right).
	\end{eqnarray}
	Here, $\tilde d_i^{\dagger}$, $\tilde d_i$ are the fermionic creation/annihilation operators in the new basis. The new energies $\tilde{\epsilon}_{1/2}$ and couplings to the leads $\tilde V_{k1/2}$ are given by
	\begin{eqnarray}
	 	\tilde \epsilon_{1/2} &=& \frac{\epsilon_1+\epsilon_2}{2} \pm \sqrt{\left(\frac{\epsilon_2-\epsilon_1}{2}\right)^2 + \Delta^2}, \label{eq:degenerate:new_energies} \label{eq:degenerate:new_eigenenergies}\\
		\tilde V_{k1/2} &=& \frac{N_{1/2}}{\tilde \epsilon_2-\tilde \epsilon_1} \left( \pm \frac{\tilde \epsilon_{2/1}-\epsilon_1}{\Delta}V_{k1} \mp V_{k2} \right), \label{eq:degenerate:new_coupling}
	\end{eqnarray}
	where the shorthand notation $N_{1/2} = \sqrt{\Delta^2 + (\epsilon_1 - \tilde{\epsilon}_{1/2})^2}$ was used. 

	For the model systems considered in Sec. \ref{sec:degenerate:nonad} and \ref{sec:degenerate:DESVIB}, $V_{k1} = \pm V_{k2}$ holds. In this case, upon increasing $\Delta$, one of the new electronic states couples stronger to the left and weaker to the right lead whereas the other electronic state couples stronger to the right and weaker to the left lead. This results in a decreasing conductivity of the model system with increasing coupling strength $\Delta$ or decreasing level spacing $|\epsilon_2-\epsilon_1|$. In the limit $\Delta \rightarrow \infty$ or $|\epsilon_2-\epsilon_1| \rightarrow 0$, the system decomposes into a state only coupling to the left and another state only coupling to the right lead, such that the current flowing through the molecule vanishes.

\subsubsection{Influence of purely nonadiabatic electronic-vibrational interactions}\label{sec:degenerate:nonad}
	We first consider transport through a molecular system exhibiting \mbox{(quasi-)}degenerate electronic states with a purely nonadiabatic electronic-vibrational coupling. This scenario is
	described by model system DESNONAD with parameters specified in Tab.\ \ref{tab:parameters}. 
	It is noted that model DESNONAD is analogous to the model system DES studied by H\"artle et.\ al.\ \cite{Haertle13} but extended by the nonadiabatic electronic-vibrational coupling. The system without electronic-vibrational coupling is known to exhibit strong destructive interference effects which suppress the current. Therefore, the model system allows for a systematic investigation of the influence of purely nonadiabatic electronic-vibrational interactions on interferences.

	\begin{figure}[tb]
		\hspace{-0.5cm}
 		\includegraphics[width=0.5\textwidth]{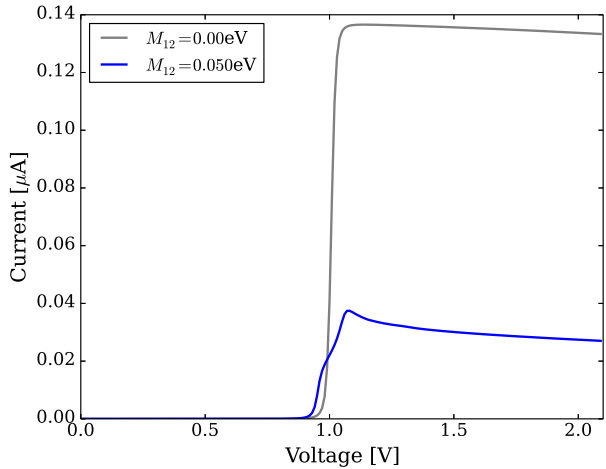}
		\caption{Current-voltage characteristics of the system DESNONAD for $\epsilon_2 = 0.505$eV with and without nonadiabatic electronic-vibrational coupling.}
		\label{fig:degenerate:I-V-DESNONAD}
	\end{figure}
	\begin{figure}[tb]
		\hspace{-0.5cm}
 		\includegraphics[width=0.5\textwidth]{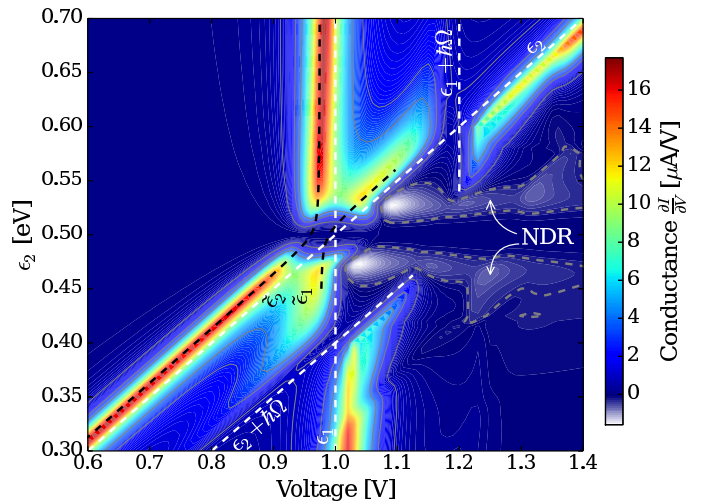}
		\caption{Conductance map of the system DESNONAD for varying electronic energy $\epsilon_2$. The purely nonadiabatic electronic-vibrational coupling strength is $M_{12} = 0.05$eV. The white dashed lines indicate the position of the electronic resonances. The positions of the corresponding two-level system with empirically determined constant coupling strength $\Delta=0.0075$eV is marked by black dashed lines in the relevant region. The light colors on the right hand side of the plot close to the line $\epsilon_2 = \epsilon_1=0.5$eV indicate regions of negative differential resistance.}
		\label{fig:degenerate:CMap-DESNONAD}
	\end{figure}
	An example of a typical current-voltage characteristic for system DESNONAD with electronic energy $\epsilon_2 = 0.505$eV is given in Fig.\ \ref{fig:degenerate:I-V-DESNONAD}, where the current for the system with nonadiabatic electronic-vibrational coupling $M_{12} = 0.05$eV is compared to that without electronic-vibrational coupling.
	The energy difference between the electronic states plays a crucial role for the interference effects. A more systematic investigation of the model system DESVIB can therefore be obtained by studying the transport properties as a function of the electronic energy $\epsilon_2$ while keeping all other parameters fixed as is done in the conductance map in Fig.\ \ref{fig:degenerate:CMap-DESNONAD}.

	Fig.\ \ref{fig:degenerate:I-V-DESNONAD} shows that the current for the interacting system is significantly smaller than that for the noninteracting system. Furthermore, it exhibits two resonance steps instead of one. The more detailed conductance map shows two distinguishable resonances for any value of $\epsilon_2$, and an  avoided level crossing as well as vanishing conductance for $\epsilon_2 = \epsilon_1$. Furthermore, negative differential resistances (NDR), that is a decrease in the current upon an increase in the bias voltage, is observed close to the line $\epsilon_2 = \epsilon_1$ for bias voltages when the second electronic resonance enters the bias window. Apart from that, the conductance map shows two additional avoided level crossing for $\epsilon_2 + \hbar\Omega = \epsilon_1$ and $\epsilon_1 + \hbar\Omega = \epsilon_2$ (intersection of white dashed lines). This last effect has already been discussed in Sec.\ \ref{sec:normal:ad_vs_nonad} and will not be further addressed here.

	To rationalize these findings, we consider the simplified noninteracting model system with constant interstate coupling from Sec.\ \ref{sec:degenerate:constant_coupling}.
 	A coupling between the electronic states of the molecule leads to a renormalization of the electronic energies and therefore to an avoided level crossing, resulting in two steps in the corresponding current. To a good approximation, the location of the electronic resonances in the conductance map of the interacting system can be described by the noninteracting model with an effective coupling of $\Delta = 0.0075$ eV (black dashed lines in Fig.\ \ref{fig:degenerate:CMap-DESNONAD}).

	In addition to the level splitting, the interaction between the molecular electronic states leads to new eigenstates.
	According to Eq.\ (\ref{eq:degenerate:new_coupling}), one of the new eigenstates couples stronger to the left, but weaker to the right lead, whereas the other eigenstate couples stronger to the right, but weaker to the left lead, which results in an overall suppression of the current.
	In the limit $\epsilon_1 = \epsilon_2$, the two new eigenstates couple either to the left or the right lead such that the current vanishes.
	This behavior is also found in the noninteracting system without interstate coupling ($\Delta = 0$) for $\epsilon_1 = \epsilon_2$. In this case, however, the current disappears due to perfect destructive interference between the two transport channels through the individual electronic states. 
	Assuming that the nonadiabatic electronic-vibrational coupling provided a source of decoherence, as was found for adiabatic electronic-vibrational coupling \cite{Haertle13}, the perfect destructive interference would be diminished leading to a finite current for $\epsilon_1 = \epsilon_2$. This is, however, not observed in the data presented in Fig.\ \ref{fig:degenerate:CMap-DESNONAD}.
	This is due to the fact that, in contrast to adiabatic electronic-vibrational coupling, nonadiabatic coupling does not provide information on the specific path (electronic state) an electron has taken through the molecular bridge and thus does not act as a direct source of decoherence.  Rather it mixes the two electronic states. This is discussed in more detail in Sec.\ \ref{sec:degenerate:DESVIB}.

	The NDR phenomenon observed in Fig.\ \ref{fig:degenerate:CMap-DESNONAD} cannot be explained by the simple toy model of Sec.\ \ref{sec:degenerate:constant_coupling} with constant coupling $\Delta$ between the two electronic states. The nonadiabatic interaction between the electronic and vibrational degrees of freedom results in an effective interstate coupling, which depends on the population of the electronic states and, thus, on the bias voltage.
	Specifically, in model system DESNONAD an increase in the population of the electronic states, obtained for bias voltages beyond the onset of resonant transport, leads effectively to a stronger interstate coupling. The resulting enhanced mixing of the electronic levels causes a decreased current, i.e.\ NDR. As this mechanism is less effective for larger energy differences $|\epsilon_2 -\epsilon_1|$, NDR can only be observed for parameter regimes close to $\epsilon_1 =\epsilon_2$.

\subsubsection{Influence of adiabatic and nonadiabatic electronic-vibrational interactions}\label{sec:degenerate:DESVIB}

	Finally, we study the influence of nonadiabatic electronic-vibrational coupling on a \mbox{(quasi-)}degenerate electronic system in the presence of adiabatic vibrational interaction. To this end, we extend the model system DESVIB introduced by H\"artle et.\ al.\ \cite{Haertle13} by a nonadiabatic electronic-vibrational coupling of varying strength (cf.\ Tab.\ \ref{tab:parameters}). 
	\begin{figure}[tb]
		\hspace{-0.5cm}
 		\includegraphics[width=0.5\textwidth]{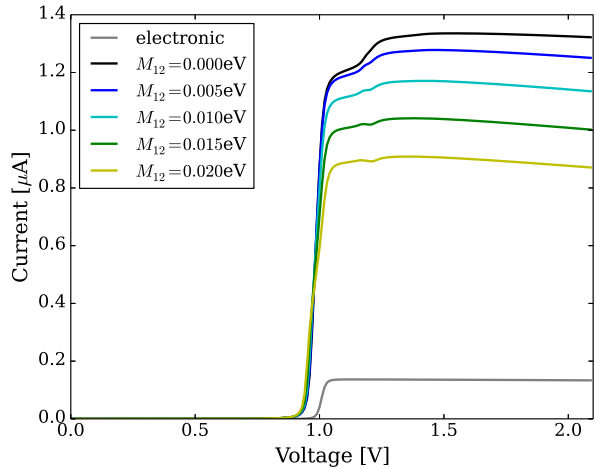}
		\caption{Current-voltage characteristics of the system DESVIB for varying nonadiabatic electronic-vibrational coupling. The grey line gives the current without electronic-vibrational interaction, the black line is the current for the system with pureley adiabatic electronic-vibrational coupling and the blue and green lines corresponds to the current of the system with different nonadiabatic electronic-vibrational coupling ranging from $M_{12} = 0.005$eV to $M_{12} = 0.02$eV.}
		\label{fig:degenerate:I-V-DESVIB}
	\end{figure}

	The results obtained for the current-voltage characteristics of  this model, depicted Fig.\ \ref{fig:degenerate:I-V-DESVIB}, show that the system with electronic-vibrational interaction allows for a significantly higher current than the purely electronic model. Thereby, the system with only adiabatic coupling to the vibrations ($M_{12} = 0$) is the best conductor, whereas the current decreases with increasing nonadiabatic electronic-vibrational coupling strength $M_{12}$.
	The overall shape of the current-voltage characteristics is rather insensitive to the strength of nonadiabatic coupling. The minor changes in the overall shape  of the current can be traced back to the interplay between adiabatic and nonadiabatic electronic-vibrational interactions as already discussed in Sec.\ \ref{sec:normal:interpaly}.

	As was studied in detail in Refs.\ \cite{Haertle11_3,Haertle13},  the interaction between electrons and nuclei   increases the current in the system with purely adiabatic coupling ($M_{12} = 0$), because the vibrations provide which-path information thus quenching the destructive interference present in the noninteracting system. \cite{Haertle13}. 
	\begin{figure}[tb]
		\hspace{-0.5cm}
 		\includegraphics[width=0.5\textwidth]{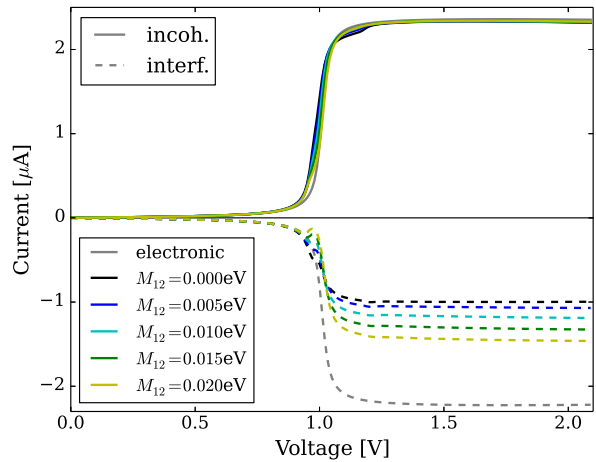}
		\caption{Interference (dashed lines) and incoherent (solid lines) contributions to the currents of system DESVIB for varying nonadiabatic electronic-vibrational coupling strength $M_{12}$.}
		\label{fig:degenerate:I-V-DESVIB:identifications}
	\end{figure}
	To analyze the influence of nonadiabatic electronic-vibrational coupling, we separate the current into incoherent and interference contributions according to Eq.\ (\ref{eq:transmission_incoherent}) and (\ref{eq:transmission_interference}). The results in Fig.\ \ref{fig:degenerate:I-V-DESVIB:identifications} show that nonadiabatic coupling reduces the interference contribution to the current, while the incoherent contribution is essentially unaffected. 
	This can be rationalized in the following way. State-specific adiabatic electronic-vibrational coupling provides information on the specific path (electronic state) an electron has taken through the molecular bridge and thus acts as a direct source of decoherence. Nonadiabatic electronic-vibrational coupling, on the other hand, mixes the two electronic states and thus tends to quench which-path information. As a result, in a system with strong decoherence such as model DESVIB, the current decreases with increasing nonadiabatic coupling.

\section{Conclusions}
\label{sec:conclusions}

	We have investigated the influence of nonadiabatic electronic-vibrational interactions on the transport properties of single-molecule junctions.
	Employing nonequilibrium Green's functions within the self-consistent Born approximation to simulate the current-voltage characterisits, we studied model molecular junctions comprising two electronic states coupled to a vibrational mode. Thereby, two different types of molecular junctions were considered, which differ by the energy spacing of the electronic states.

	For molecular junctions with well separated electronic states, an analytical analysis of the most important vibrational effects shows that nonadiabatic coupling may have pronounced effects if the resonance condition $\epsilon_i + \hbar\Omega \approx \epsilon_j$ with $i\neq j$ is fulfilled. 
	Numerical calculations revealed the fundamental differences between adiabatic and nonadiabatic electronic-vibrational coupling, which manifest themselves in a different dependence of the transport properties on the vibrational energy $\hbar\Omega$ and in transport channels caused by the level-mixing in the case of nonadiabatic coupling. The latter mechanism leads to an asymmetric current-voltage characteristic, i.e.\ rectifying behavior, in junctions with state-specific asymmetric coupling to the leads. Moreover, the interplay between adiabatic and nonadiabatic vibrational effects results in interference effects that change the current depending on the relative sign between the different vibrational couplings. 

	To analyze the effect of nonadiabatic electronic-vibrational interactions in molecular junctions with \mbox{(quasi-)}degenerate electronic levels, we considered model systems, which were previously shown to exhibit significant  interference and decoherence effects \cite{Haertle13}.
	Our study shows that purely nonadiabatic vibrational coupling in these systems results in an avoided crossing of the energy levels and tends to suppress the current. 
	Moreover, the specific nature of the nonadiabatic electronic-vibrational interaction can result in NDR as a consequence of charging the molecule.
	In contrast to state-specific adiabatic coupling, nonadiabatic coupling does not provide a decoherence mechanism. Rather, it causes mixing of the electronic states and thus quenches which-path information.   
	
	In the present work we have considered molecular junctions with a single vibrational mode. In polyatomic molecules, multiple vibrational modes may interact with the electronic degrees of freedom. In this case intersections of potential energy surfaces are possible and may result in strong nonadiabatic coupling. The resulting phenomena are well studied in molecules in the gas phase and in solution \cite{Domcke04}. The investigation of these effects under nonequilibrium conditions in molecular junction is an interesting topic for future research.

\section*{Acknowledgments} 
	This work has been supported by the Deutsche
	Forschungsgemeinschaft (DFG) through the DFG-Cluster of
	Excellence 'Engineering of Advanced Materials', SFB 953,
	and a research grant. The generous allocation
	of computing time by the computing centers in Erlangen
	(RRZE) and Munich (LRZ) is gratefully acknowledged.
	RH was supported by the Alexander von Humdoldt foundation via a Feodor-Lynen research fellowship.

\appendix
\section{Identification of important transport processes for systems with well separated electronic states}\label{appendix:normal:identification}

	In Sec.\ \ref{sec:normal:identification}, we identified the most important effects of electronic-vibrational interactions by expanding the current in the couplings $M_{ij}$ and $V_{ij}$.
	Considering well separated electronic systems and based on the strongly peaked nature of Green's functions, we employed an additional approximation beyond perturbation theory in Sec.\ \ref{sec:normal:identification}. Specifically, a simplification for the expression for the current was obtained by neglecting the product of two Green's functions of different electronic states for a fixed energy. 
	Depending on the system under consideration, the coupling to the leads can result in relatively large off-diagonal Green's function elements such that completely disregarding the product of two Green's functions of different electronic states at the same energy may not be valid.
	Nevertheless, the coupling to the leads still represents a perturbation to the system such that an improved expression for the current can be derived by additionally allowing for contributions that contain the product with one Green's function of another electronic state at a given energy. A generalization of this approach is straightforward. The elastic correction to the current and the inelastic current read
	\begin{widetext}
	\begin{eqnarray}
	\delta  I_{\text{L el}} &\approx&  
			\frac{e}{2\pi\hbar} \sum_{i, j\neq i} \int 
			  \Sigma_{\text{L }ij}^<(\epsilon) \left( G_{jj}^{0r}(\epsilon)^2 \left[ |M_{ji}|^2 \mathcal{G}_{ii}^{0r}(\epsilon) + M_{jj}^2 \mathcal{G}_{jj}^{0r}(\epsilon) + \sum_{a\neq i\land j} |M_{ja}|^2 \mathcal{G}_{aa}^{0r}(\epsilon) \right] G_{ii}^{0a}(\epsilon) + \text{h.c.} \right) \Sigma_{\text{R }ji}^>(\epsilon)  \nonumber \\ &&
			+ \Sigma_{\text{L }ii}^<(\epsilon) \left( |G_{ii}^{0r}(\epsilon)|^2 \left[ M_{ii} M_{ij} \mathcal{G}_{ii}^{0r}(\epsilon) + M_{ij} M_{jj} \mathcal{G}_{jj}^{0r}(\epsilon) + \sum_{a\neq i\land j} M_{ia} M_{aj} \mathcal{G}_{aa}^{0r}(\epsilon) \right] G_{jj}^{0r}(\epsilon) + \text{h.c.} \right) \Sigma_{\text{R }ji}^>(\epsilon) \nonumber \\ &&
			+ \Sigma_{\text{L }ij}^<(\epsilon) \left( G_{jj}^{0r}(\epsilon) \left[ M_{ji} M_{ii} \mathcal{G}_{ii}^{0r}(\epsilon) + M_{jj} M_{ji} \mathcal{G}_{jj}^{0r}(\epsilon) + \sum_{a\neq i\land j} M_{ja} M_{ai} \mathcal{G}_{aa}^{0r}(\epsilon) \right] |G_{ii}^{0r}(\epsilon)|^2 + \text{h.c.} \right) \Sigma_{\text{R }ii}^>(\epsilon)
			\text{d}\epsilon \nonumber \\  &&
			+\delta I_{\text{L el}}^{\text{sep}} + \text{Renorm.} + \mathcal{O}(M^4 V^6), \label{eq:appendix:1}
	\end{eqnarray}
	\begin{eqnarray}
	 I_{\text{L inel}} &\approx& I_{\text{L inel}}^{\text{sep}} +
			\frac{e}{2\pi\hbar} \int 
			  \sum_{i,a\neq i} \Sigma_{\text{L }ii}^<(\epsilon) |G_{ii}^{0 r}(\epsilon)|^2 \left[ M_{ii} M_{ai} \mathcal{G}_{ia}^{0>}(\epsilon) 
			+ M_{ia} M_{ii} \mathcal{G}_{ai}^{0>}(\epsilon) + \sum_{b\neq a\land i} M_{ia} M_{bi} \mathcal{G}_{ab}^{0>}(\epsilon)\right] \nonumber \\ &&
			+ \sum_{i ,j\neq i} \Sigma_{\text{L }ij}^<(\epsilon) G_{jj}^{0 r}(\epsilon) \left[ M_{ji} M_{ii} \mathcal{G}_{ii}^{0>}(\epsilon)  
			+ M_{jj} M_{ji} \mathcal{G}_{jj}^{0>}(\epsilon)  + \sum_{a\neq i\land j} M_{ja} M_{ai} \mathcal{G}_{aa}^{0>}(\epsilon) \right] G_{ii}^{0 a}(\epsilon) 
			\text{d}\epsilon + \mathcal{O}(M^4 V^6). \label{eq:appendix:2} \nonumber \\
	\end{eqnarray}
	\end{widetext}
	Thereby, $\delta I_{\text{L el}}^{\text{sep}}$ and $I_{\text{L inel}}^{\text{sep}}$ are the contributions to the current as specified in Eq.\ (\ref{eq:identification:el_corr}) and (\ref{eq:identification:inel}), Renorm.\ denotes the influence of the Hartree term which leads to further renormalization of the electronic levels and which will be neglected here for the sake of readability. In this approximation, the expressions for the currents incorporate a variety of different terms. The equations can be separated into three different parts corresponding to different type of effects of the vibrations on the current. 

	The first effect is an enhancement of purely adiabatic or nonadiabatic vibrational effects identified in Sec. \ref{sec:normal:identification}. These terms are proportional to $M_{ii}^2$ and $|M_{ij}|^2$ with $i\neq j$ respectively. 
	The second type mixes the adiabatic and the nonadiabatic electronic-vibrational coupling. The corresponding terms are proportional to $M_{ii} M_{ij}$ with $i\neq j$. Notice that these summands are also proportional to off-diagonal elements of the self-energy for the coupling to the leads. As discussed in Sec. \ref{sec:normal:interpaly}, this combined action of adiabatic and nonadiabatic electronic-vibrational interaction can have a strong influence on the current. 
	The last kind of processes that influences the current in this approximation depends on three different electronic states. The corresponding terms are proportional to $M_{ia} M_{aj}$ with $i\neq j\neq a$ and rely only on nonadiabatic electronic-vibrational interactions. They describe transport processes, where an electron is scattered from an initial to a final electronic state via another, different state. They represent therefore a more general form of the purely nonadiabatic electronic-vibrational processes identified in Sec.\ \ref{sec:normal:identification}. While the latter type of processes do not contribute in our models with two electronic levels, they may be of importance in molecular junctions with multiple closely lying electronic states.

\bibliography{PaperNonAd}

\end{document}